\input harvmac\skip0=\baselineskip
\input epsf

\newcount\figno
\figno=0
\def\fig#1#2#3{
\par\begingroup\parindent=0pt\leftskip=1cm\rightskip=1cm\parindent=0pt
\baselineskip=11pt \global\advance\figno by 1 \midinsert
\epsfxsize=#3 \centerline{\epsfbox{#2}} \vskip 12pt {\bf Fig.\
\the\figno: } #1\par
\endinsert\endgroup\par
}
\def\figlabel#1{\xdef#1{\the\figno}}
\def\encadremath#1{\vbox{\hrule\hbox{\vrule\kern8pt\vbox{\kern8pt
\hbox{$\displaystyle #1$}\kern8pt} \kern8pt\vrule}\hrule}}

\def\p{\partial}

% approximately less than

% approximately greater than

%\SimonsNM
\lref\ssty{

  A.~Simons, A.~Strominger, D.~M.~Thompson and X.~Yin,
  ``Supersymmetric branes in AdS(2) x S**2 x CY(3),''
  Phys.\ Rev.\ D {\bf 71}, 066008 (2005)
  [arXiv:hep-th/0406121].
  %%CITATION = HEP-TH 0406121;%%
  %%Cited 7 times in SPIRES-HEP
}

\lref\gvtwo{
  R.~Gopakumar and C.~Vafa,
  ``M-theory and topological strings. II,''
  arXiv:hep-th/9812127.
  %%CITATION = HEP-TH 9812127;%%
  %%Cited 97 times in SPIRES-HEP
}
%\GopakumarII
\lref\gvone{
 R.~Gopakumar and C.~Vafa,
``M-theory and topological strings. I,''
arXiv:hep-th/9809187.
%%CITATION = HEP-TH 9809187;%%
}

\lref\joe{
  J.~Polchinski,
  ``String theory. Vol. 2: Superstring theory and beyond,''
  Cambridge, UK: Univ. Pr. (1998).
%\href{http://www.slac.stanford.edu/spires/find/hep/www?irn=4634802}{SPIRES entry}
}

\lref\cgms{
  M.~Cyrier, M.~Guica, D.~Mateos and A.~Strominger,
  ``Microscopic entropy of the black ring,''
  Phys.\ Rev.\ Lett.\  {\bf 94}, 191601 (2005)
  [arXiv:hep-th/0411187].
  %%CITATION = HEP-TH 0411187;%%
  %%Cited 19 times in SPIRES-HEP
}

%\VafaGR
\lref\VafaGR{
C.~Vafa,
``Black holes and Calabi-Yau threefolds,''
Adv.\ Theor.\ Math.\ Phys.\  {\bf 2}, 207 (1998)
[arXiv:hep-th/9711067].
%%CITATION = HEP-TH 9711067;%%
}
\lref\osv{
  H.~Ooguri, A.~Strominger and C.~Vafa,
  ``Black hole attractors and the topological string,''
  Phys.\ Rev.\ D {\bf 70}, 106007 (2004)
  [arXiv:hep-th/0405146].
  %%CITATION = HEP-TH 0405146;%%
  %%Cited 89 times in SPIRES-HEP
}

\lref\msw{  J.~M.~Maldacena, A.~Strominger and E.~Witten,
  ``Black hole entropy in M-theory,''
  JHEP {\bf 9712}, 002 (1997)
  [arXiv:hep-th/9711053].
  %%CITATION = HEP-TH 9711053;%%
  %%Cited 96 times in SPIRES-HEP
}
\lref\SimonsNM{ A.~Simons, A.~Strominger,
D.~M.~Thompson and X.~Yin, ``Supersymmetric branes in AdS(2) x
S**2 x CY(3),'' arXiv:hep-th/0406121.
%%CITATION = HEP-TH 0406121;%%
}
\lref\bu{
  R.~Dijkgraaf, R.~Gopakumar, H.~Ooguri and C.~Vafa,
  ``Baby universes in string theory,''
  arXiv:hep-th/0504221.
  %%CITATION = HEP-TH 0504221;%%
  %%Cited 23 times in SPIRES-HEP
}
\lref\mina{
  M.~Aganagic, H.~Ooguri, N.~Saulina and C.~Vafa,
  ``Black holes, q-deformed 2d Yang-Mills, and non-perturbative topological
  strings,''
  Nucl.\ Phys.\ B {\bf 715}, 304 (2005)
  [arXiv:hep-th/0411280].
  %%CITATION = HEP-TH 0411280;%%
  %%Cited 34 times in SPIRES-HEP
}
\lref\db{
  J.~de Boer,
  ``Six-dimensional supergravity on S**3 x AdS(3) and 2d conformal field
  theory,''
  Nucl.\ Phys.\ B {\bf 548}, 139 (1999)
  [arXiv:hep-th/9806104].
  %%CITATION = HEP-TH 9806104;%%
  %%Cited 112 times in SPIRES-HEP
}
\lref\denef{F. Denef, Talk given at Harvard Workshop on Black Holes and Topological Strings, Jan. 31 '06.}

\lref\cv{C.~Vafa,
  ``Two dimensional Yang-Mills, black holes and topological strings,''
  arXiv:hep-th/0406058.}

  \lref\gravmul{
  A.~Fujii, R.~Kemmoku and S.~Mizoguchi,
  ``D = 5 simple supergravity on AdS(3) x S(2) and N = 4 superconformal  field
  theory,''
  Nucl.\ Phys.\ B {\bf 574}, 691 (2000)
  [arXiv:hep-th/9811147].
  %%CITATION = HEP-TH 9811147;%%
  %%Cited 11 time in SPIRES-HEP
}

\def\Im{{\rm Im~}}
%\draft

\Title{\vbox{\baselineskip12pt\hbox{}
}}
{From AdS$_3$/CFT$_2$ to Black Holes/Topological Strings}
%{ M-derivation of $Z_{\rm BH}=|Z_{\rm top}|^2$}

\centerline{Davide Gaiotto,~ Andrew Strominger and Xi Yin }
\smallskip
\centerline{Jefferson Physical Laboratory, Harvard University,
Cambridge, MA 02138} \vskip .6in \centerline{\bf Abstract} {
The elliptic genus $Z_{\rm BH}$ of a large class of 4D black holes can be expressed as an M-theory partition function on an $AdS_3\times S^2\times CY_3$ attractor.
We approximate this partition function  by summing over multiparticle
chiral primary states of membranes which wrap curves in the $CY_3$ and
tile Landau levels on the horizon $S^2$.  Significantly, membranes and
antimembranes  can preserve the same supercharges if they occupy antipodal points on the horizon.  It is shown the membrane contribution to $Z_{\rm BH}$ gives precisely the
topological string partition function $Z_{\rm top}$ while the antimembranes
give $\bar Z_{\rm top}$, implying $Z_{\rm BH}=|Z_{\rm top}|^2$ in this approximation.
 } \vskip .3in

%\smallskip
\Date{}

\listtoc \writetoc
\newsec{Introduction}

A four-dimensional black hole in an M-theory  compactification on a Calabi-Yau threefold $X$ times an $S^1$ can be constructed by wrapping an M5-brane with fluxes and $S^1$ momentum on a 4-cycle $P$ in $X$.
 This black hole  is dual to the R sector of the $(0,4)$ CFT which lives on the dimensionally reduced M5 worldvolume \msw. The NS sector of this same $(0,4)$ CFT is dual to supergravity on $AdS_3\times S^2\times X$ (as well as  the
5D black ring \cgms).

The elliptic genus $Z_{BH}=Z_{CFT}$ of the 4D black hole is\foot{This is related by spectral flow to the Ramond sector trace. If the center of mass multiplet is included, there should be an extra insertion of
$F^2/\tau_2$, but we will suppress this herein. }
\eqn\egf{Z_{BH}={\rm Tr} (-)^F q^{L_0-{c_L\over 24}}e^{2\pi i q_Ay^A},}
where the trace is over chiral primary states on $AdS_3\times S^2\times X$,
$q_A$ is a membrane charge and $y^A$ the conjugate potential. We work in the dilute gas
expansion in which \egf\ is dominated by multi-particle chiral primaries states of membranes
wrapping holomorphic curves in $X$.

A crucial point in the following is that both membrane and anti-membrane
states
contribute to \egf. This is because  a membrane wrapping a holomorphic
curve $Q=q_A\alpha^A$ in $X$ and sitting at say the north pole of the $S^2$ preserves the the $same$ set of supersymmetries as the  oppositely-charged  $anti$-M2-brane wrapping $Q$ and sitting at the south pole \ssty. This may sound strange as we are used to the idea in flat space  that
branes and antibranes preserve opposite supersymmetries because they have opposite charges. However in $AdS_2\times S^2$ the $S^2$ angular momentum plays the role of the central charge in stabilizing BPS states.
Static wrapped branes
in this background carry this angular momentum much like a static electron in the field of a monopole.  Hence branes and antibranes at
anitpodal points can carry the same angular momentum and preserve the same supersymmetries.

In this paper we work out in detail the degeneracies of chiral primary wrapped membranes of all stripes and their contribution to $Z_{BH}$.
We find the product of two  complex conjugate factors, one from branes and another from antibranes. Including an additional factor from massless supergravity modes, a modular transformation factor and
using the Gopakumar-Vafa relation
\gvtwo\ between BPS degneracies and Gromov-Witten invariants,   we recover precisely the OSV relation \osv\foot{The agreement is up to factors
which depend only on $q$ and not any of the Calabi-Yau data.}
\eqn\rrf{ Z_{\rm BH}=|Z_{\rm top}|^2.}
In \osv, this perturbative factorization was discovered by direct brute force
computation. In the present work we have found a simple physical explanation:
$Z_{\rm top}$ is the membrane contribution, while $\bar Z_{\rm top}$ is the anti-membrane contribution.\foot{Related discussions of this factorization appear in \refs{\cv\mina \bu-\denef} .}We further hope that the the framework can be used to
systematically compute non-perturbative corrections to \rrf, such as perhaps arising from chiral primary wrapped  M5 branes.

This paper is organized as follows. Section 2
recaps the M-theory attractor $AdS_3\times S^2\times X$.
The classical and quantum BPS states of wrapped M2-branes
are described in section 3. In section 4 we compute the elliptic genus
from the bulk theory, including the contribution
from wrapped M2-branes, massless bulk supergravity fields and boundary singletons. The
result is compared to the black hole partition function and
topological strings in section 5. Appendix A reproduces
a useful resummation formula involving the Gopakumar-Vafa invariants.

\newsec{Preliminaries}
We consider M-theory on an $AdS_3\times S^2\times X$ attractor geometry, where
$X$ is a Calabi-Yau threefold, with 4-form flux
\eqn\fform{ G_4 = \omega_{S^2} \wedge  p^A \omega_A }
Here $\omega_A$ is a basis of harmonic 2-forms dual to 2-cycles
$\alpha^A$ in $X$ with intersection  numbers $\int_X\omega_A\wedge \omega_B\wedge \omega_C=6D_{ABC}$. The metric is :\foot{We adopt 11D Planck units in which, as in \joe,  the action is
$(2\pi)^{-8}\int d^{11}x\sqrt{-g}R+\cdots$.}
\eqn\volk{ \int_{\alpha^A} J = ( 2\pi)^2 {p^A\over {\ell}} }
\eqn\met{ds_{11}^2= \ell^2(-\cosh^2\chi d\tau^2+d\chi^2+\sinh^2\chi d\psi^2)+
{\ell^2 \over 4}(d\theta^2+\sin^2 \theta d\phi^2)+ds_X^2,}
where ${\ell}$ is the radius of $AdS_3$ and  $J$ is the   K\"ahler form on $X$ .  This is the near horizon attractor geometry of an M5-brane
wrapped on the 4-cycles $p^A \Sigma_A$ (where  $\Sigma_A$ are a basis of 4-cycles dual to $\omega_A$) in $X$ and forming an extended string in the noncompact five dimensions.
One
expects M-theory on $AdS_3\times S^2\times X$ to be dual to the
$(0,4)$ CFT on the
 M5-brane world volume dimensionally reduced on $P$.
This CFT has leading order left central charge $c_L={3\ell \over 2G_3}=6D_{ABC}p^Ap^Bp^C$,
where the 3D Newton constant is $G_3= {16\pi^7 \over Vol_XVol_{S^2}}$
\msw.

\newsec{BPS wrapped branes}
\subsec{Classical}The geometry \met\ has a classical supersymmetric
M2-brane wrapping a holomorphic genus $g$ curve in the class  $C=q_A\alpha^A$ and sitting at the center of $AdS_3$, $\chi=0$. The kappa-symmetry analysis is very similar to the analysis of D2-branes in
\ssty\ and will not be repeated here.
It can sit at any point on the $S^2$, but the unbroken supersymmetries
vary as the point moves on the $S^2$. We will take it to sit at the north pole with $\theta=0$. There is also an oppositely charged configuration consisting of an
anti-membrane at the south pole ($\theta=\pi$) which preserves the same supersymmetries.

Both of these configurations have non zero angular momentum corresponding to $\phi$ rotations of the  $S^2$ and
given by
\eqn\wag{J^3=\half  q_Ap^A.}
This angular momentum is carried by the fields much as for a monopole-electron pair in 4D. Although the M2 and anti-M2 have opposite charges, they still carry the same sign  $J^3$ because they sit at opposite poles. Since they are static and saturate
the BPS bound $L_0=J^3$, it follows that $AdS_3 $ mass and angular momentum are classically
\eqn\diu{L_0=\overline L_0 = \half q_Ap^A.}
This agrees with a direct calculation of the mass $M={\ell }(L_0+\bar L_0)$ as the membrane tension $1 \over( 2 \pi)^2$ times the membrane area
${(2\pi)^2 \over \ell} q_Ap^A$.

\subsec{Quantum}

Quantum mechanically, the M2 fluctuates over the moduli space ${\cal
M}_C$ of the genus $g$ curve $C$ in $X$, and has a degeneracy from
worldvolume fermion zero modes. The supersymmetric quantum ground
states correspond to cohomology classes on ${\cal M}_C$ and BPS
hypermultiplets in 5D. This problem was studied in the context of
compactification to 5D Minkowski space in   \gvtwo, where the
hypermuliplets have $SO(4)\sim SU(2)_L\times SU(2)_R$ spin
content
 \eqn\rtop{\sum N_{Q,j_L,j_R}\bigl( [(0,\half)\oplus 2
(0,0)]  \otimes(j_L,j_R) \oplus [(\half,0)] \oplus 2(0,0)]\otimes(j_R,j_L) \bigr) } for some integers
$N_{Q,j_L,j_R}$ which depend on $X$ and $Q$, the homology class of
$C$. The range of $j_L$ is determined by the genus of $C$, and $j_R$
is related to the weight of Lefschetz action on the moduli space of
$C$ \gvtwo.

 We wish to find the supersymmetric ground states - i.e. chiral primaries-  of these hypermultiplets on
 $AdS_3\times S^2$. In this case the unbroken global superalgebra is
 $SU(1,1|2)$ and the relevant central charge is the angular momentum  $J^3$ rather
 than the (graviphoton component of the) charge $q_A$ (this problem was considered \db\gravmul ).
Due to its coupling to the 4-form flux \fform, the M2-brane feels a
magnetic field on the $S^2$ of $B=q_Ap^A$ units of flux. This leads
to Landau levels on the $S^2$ which fall into representations of the
$SU(2) \in SU(1,1|2)$ rotation.
The highest weight states arising
from a hypermulitplet in the representaion $(j_L,j_R)$ have total
spin
\eqn\hiu{J^3= \half q_Ap^A +m_R+m_L+\half+l,}  where  $l\geq 0,~-j_{L,R}\le m_{L,R} \le j_{L,R}$, and $l$ is the
orbital angular momentum on the $S^2$.
 $m_R+m_L$ appears in this expression because the $U(1)\in SU(2)$ rotations of $S^2$ correspond to
a diagonal $U(1)$ rotation in the $SU(2)_L\times SU(2)_R $ of $R^4$. The shift of $\half$ appears because of the tensor product with a spin half hypermultiplet appearing in the definition \rtop.

The BPS chiral primary bound implies that these states have $\bar
L_0=J^3$. They are multiplets under the $SL(2,R)_L$ conformal
algebra which commutes with $SU(1,1|2)$ and acts on the $AdS_3$
component of the wavefunction.   There is one lowest weight state
with $ L_0 =\bar L_0+m_L-m_R+\half=\half q_A p^A+2m_L+l+1$ for
$-j_L\le m_L \le j_L,~~-j_R\le m_R \le j_R$. $m_L-m_R$ appears in this expression because the $U(1))$ spatial rotations of $AdS_3$ correspond to
an  anti-diagonal $U(1)$ rotation in the $SU(2)_L\times SU(2)_R $ of $R^4$. Each of these has a further tower of chiral
primary descendants
obtained by acting with $L_{-1}$.

In summary for every charge $q_A$ $(j_L,j_R)$ hypermultiplet there is one chiral primary with
\eqn\opx{L_0=\half q_A p^A+2m_L+l+1+J_\phi}
for every integrally-spaced value of
\eqn\dvz{-j_L\le m_L \le j_L,~~-j_R\le m_R \le j_R,~~l \ge 0, ~~J_\phi \ge 0.}

In addition there are antimembrane chiral primaries.
M-theory with no branes is invariant under parity $P$, which we take to
interchange the north and south pole of the $S^2$ .
When branes are added it is invariant under $CP$ where
$C$ reverses the brane charges.  $L_0$ and $J^3$ are $CP$ invariant.
Hence the action of $CP$ on a chiral primary gives another  chiral primary.
In the case at hand it turns each of the above M2-brane states into an antipodally located anti-M2-brane state.
However these states will
contribute differently to the elliptic genus because they have different charges.

\newsec{The elliptic genus  on $AdS_3\times S^2 \times X$ }
In this section we compute the supergravity elliptic  genus from  M theory.
In the NS sector this is given by\foot{Note that we are not including the factor
of $q^{-{c_L\over 24}}$ in $Z_{sugra}$, and the $L_0$ entering here arises only the wrapped branes.  This factor corresponds to the ground state energy of $AdS_3$. }
 \eqn\deft{Z_{sugra} (\tau,
y^A)  = {\rm Tr}_{\overline L_0 = J_R^3} (-)^F e^{2\pi i\tau
L_0 +
2\pi i q_A y^A}}
We work in the dilute gas  approximation  in which the density of chiral primaries is low. This is the case for large $\Im \tau $
and/or large  $\Im y^A$.
There will be two kinds of contributions, one from wrapped membranes  and one from supergravity modes, which are computed in the next two subsections.
\subsec{Wrapped membranes}
In this subsection we find the chiral primaries corresponding to membranes wrapped on holomorhpic curves in $X$.

\centerline{\bf Genus zero}
 For simplicity let's first consider the case of an
isolated rational genus zero curve with degeneracy $N_{q_A}$, so that
there is no internal $(j_L,j_R)$ contribution. Summing over multiparticle states of this variety with weights and multiplicities given in \opx,\dvz\ gives
\eqn\ellfs{ \eqalign{ Z^0_{sugra}  & = \prod_{q_A,l\ge0 ,J_\psi \ge 0} {
(1-e^{2\pi i\tau (\half q_Ap^A + l+J_\psi+1)} e^{2\pi iq_A y^A}
)^{N_{q_A}}} \cr & ~~~\times  \prod_{q_A,l\ge 0,J_\psi  \ge 0} { (1-e^{2\pi
i\tau (\half q_Ap^A + l+J_\psi+1)} e^{-2\pi iq_A y^A} )^{N_{q_A}}}  } }
The first factor comes from M2-branes while the second comes from anti-M2-branes.
We can reorganize the product by defining $n=l+J_\psi+1$
\eqn\zsud{ \eqalign{ Z^0_{sugra}(\tau,y^A) =
\prod_{q_A,n>0} { (1-e^{2\pi i\tau (\half q_Ap^A + n)}
e^{2\pi iq_A y^A} )^{nN_{q_A}}} \cr \times  \prod_{q_A,n>0}
{ (1-e^{2\pi i\tau (\half q_Ap^A + n)} e^{-2\pi iq_A y^A}
)^{nN_{q_A}}}  }}\centerline{\bf Higher genus}
For general $(j_L,j_R)$, instead of \ellfs\ we have
\eqn\llfs{ \eqalign{ Z^{j_L,j_R}_{sugra}  & = \prod_{q_A,l ,J_\psi , m_L,m_R} {
(1-e^{2\pi i\tau (\half q_Ap^A + l+J_\psi+1+2m_L)} e^{2\pi iq_A y^A}
)^{(-)^{2j_R+2j_L}N_{q_A,j_L,j_R}}} \cr & ~~~\times  \prod_{q_A,l,J_\psi , m_L,m_R} { (1-e^{2\pi
i\tau (\half q_Ap^A + l+J_\psi+1+2m_L)} e^{-2\pi iq_A y^A} )^{(-)^{2j_R+2j_L}N_{q_A,j_L,j_R}}} \cr & = \prod_{q_A,n, m_L,} {
(1-e^{2\pi i\tau (\half q_Ap^A + n+2m_L)} e^{2\pi iq_A y^A}
)^{(-)^{2j_R+2j_L}n(2j_R+1)N_{q_A,j_L,j_R}}} \cr & ~~~\times  \prod_{q_A,n, m_L} { (1-e^{2\pi
i\tau (\half q_Ap^A + n+2m_L)} e^{-2\pi iq_A y^A} )^{(-)^{2j_R+2j_L}n(2j_R+1)N_{q_A,j_L,j_R}}},  } }
where $J_\psi$ and $l$ are non-negative integers, $n$ is a positive integer  and $-j_{L,R}\le m_{L,R} \le j_{L,R}$.
We will see below that these terms give all the loop contributions of the
squared topological string partition function.
\subsec{Supergravity modes}
The massless spectrum of M-theory compactified on $X$ consists of
$n_H=2(h^{2,1}(X)+1)$ hypermultiplets, $n_V=h^{1,1}(M)-1$ vector
multiplets, and a graviton multiplet.
Their spectrum on $AdS_3\times S^2$
organizes into short representations of $SL(2,{\bf R})\times
SU(1,1|2)$. The corresponding chiral primaries can be labelled by their
$(L_0, \bar L_0=J_R^3)$ quantum numbers, with the spectrum:
\eqn\masslesp{ \eqalign{ & n_H\bigoplus_{l\geq 0}(l+1,l+\half) + n_V
\bigoplus_{l\geq 0}\left[(l+1,l+1)+(l+1,l )\right] \cr
& + \bigoplus_{l\geq 0}
\left[ (l+1,l+2)+(l+1,l+1)+(l+1,l)+(l+2,l ) \right] } }
The spectrum is obtained in \refs{\db,\gravmul}.  We have assumed here  the
range of allowed values of $l$ is so as to include all possibilities with $L_0>0$.
Whether or not singleton contributions with $\bar L_0=0$ should be included
is a subtle issue which depends on the details of the asymptotic $AdS_3$
boundary conditions, and is beyond the scope of this paper.  The ambiguity
leads to terms that depend on $q$ but not any of the Calabi-Yau data.

As before, one can act on them with $L_{-1}$ and generate
further chiral primary states
with nonzero orbital angular momenta $J_\psi$ in $AdS_3$.
The contribution from  \masslesp\
to the elliptic genus is
\eqn\ellmasl{ \eqalign{ & \prod_{l,J_\psi\geq 0}{(1-q^{l+J_\psi+1})^{n_H}
\over (1-q^{l+J_\psi+1})^{2n_V+3}
(1-q^{l+J_\psi+2}) }
\cr &= \prod_{n\geq 1}(1-q^n)    M(q)^{-\chi(X)} } }
where $M(q)=\prod_{n\geq 1}(1-q^n)^n$ is the Macmahon function and
$\chi(X)=2(h^{1,1}-h^{2,1})$ is the Euler characteristic of $X$.
We will henceforth drop the $\eta$ function prefactor which does not depend on
Calabi-Yau data.
 The net contribution from massless neutral  supergravity modes including singletons is then simply
\eqn\rtup{Z^{massless}_{sugra}=\prod_{n\geq 1}(1-q^n)^{-n\chi}.}

\subsec{Putting it all together}
Let us now summarize and compile the results of this section into a formula for $Z_{BH}=Z_{CFT}$ The elliptic genus of the $(0,4)$ CFT as a Ramond sector trace is \foot{There are some subtleties here in the spectral flow related to the
fact that the $U(1)$ current involves membrane charges and is in a supermultiplet with the center of mass degrees of freedom \msw\ which we shall not try to address. }
\eqn\elltr{ Z_{CFT}(\tau, y^A) = {\rm Tr}_R (-)^F q^{L_0-{c_L\over 24}}
\bar q^{\bar L_0-{c_R\over 24}} e^{2\pi i y^A q_A} }
In the dilute gas expansion around $\Im \tau \to \infty$
\eqn\ttw{\eqalign{Z_{CFT}(\tau, y^A)&=e^{-\pi i \tau c_L /12} Z_{sugra}(\tau, y^A)\cr
& =e^{-\pi i \tau c_L /12}\prod (1-e^{2\pi i \tau n})^{-n\chi} \cr
&  ~~~\times \prod{
(1-e^{2\pi i\tau (\half q_Ap^A + n+2m_L)} e^{2\pi iq_A y^A}
)^{(-)^{2j_R+2j_L}n(2j_R+1)N_{q_A,j_L,j_R}}} \cr  & ~~~\times  \prod { (1-e^{2\pi
i\tau (\half q_Ap^A + n+2m_L)} e^{-2\pi iq_A y^A} )^{(-)^{2j_R+2j_L}n(2j_R+1)N_{q_A,j_L,j_R}}},  } }
where we take the products over  the $q_A$ charge lattice, positive integral
$p,n$, integral or half integral $j_L,j_R$ and $-j_{L,R}\le m_{L,R} \le j_{L,R}$.

\newsec{Derivation of OSV}
In the preceding section we computed the dilute gas approximation to the elliptic genus of the
$(0,4)$ CFT, denoted $Z_{BH}$,  as a product of terms coming from massless supergravity modes and wrapped membranes. In this section we wish to compare our result
with the OSV formula \osv\ for the same object as the square of the  topological
string partition function $Z_{top}$.  The OSV result begins with the Bekenstein-Hawking relation and then includes all orders perurbative corrections. This is the regime in which many BPS excitations are present and is the opposite of a dilute gas.
However, modular invariance relates the dilute gas
to the high-temperature regime needed for comparison to OSV as follows.

Under the modular transform $\tau\to -1/\tau$, we have
\eqn\modultr{ \eqalign{ Z_{BH}=Z_{CFT}(\tau, y^A) &= Z_{CFT}(-1/\tau,y^A/\tau)
e^{-{2\pi i\over \tau}y^2 }
\cr &= \exp\left[{{2\pi i\over \tau}\left({c_L\over 24}-y^2\right)}\right]
 Z_{sugra}(-1/\tau,y^A/\tau) } }
where $y^2 = D_{ABC}p^A y^B y^C$. In this form we can consider high temperatures $\Im \tau \to 0$ since the RHS will then involve $Z_{sugra}$ at
low temperatures.
The $(0,4)$ CFT of \msw\ has $c_L=6D+c_2\cdot P = 6D_{ABC}p^Ap^Bp^C+c_{2A}p^A$,
The prefactor in \modultr\ is then
\eqn\ufact{ \exp\left[{2\pi i\over\tau}\left(
{c_L\over 24}-y^2 \right) \right] = \exp\left\{ {\pi^2\over \phi^0}
\left[ D_{ABC}p^A\left( p^Bp^C-{\phi^B\phi^C\over \pi^2}
\right)+{1\over 6}c_{2A}p^A \right] \right\} }
This is precisely $|\exp({\cal F}_0^{(0)}+{\cal F}_1^{(0)})|^2$,
where ${\cal F}^{(0)}_{0,1}$ denote the part of topological string amplitude
that is perturbative on the world sheet, at genus 0 and 1.
$Z_{sugra}$ then give the rest of $|Z_{top}|^2$.  To see this we need to use
the fundamental  relation between the integral degeneracies $N_{q_A,j_L,j_R}$ of BPS states and the coefficients of the topological string expansion found from a Schwinger computation in \gvone. Indeed comparing with
\gvone\foot{To see this agreement one needs to use a well-known resummation
of the formulae of \gvone\ which does not seem to be in the literature. For the readers benefit we reproduce this in the appendix.}  we find precisely that, for
purely imaginary $\tau=i\phi^0/2\pi$ and $y^A=i\phi^A/2\pi$,

\eqn\reexp{ \eqalign{Z_{BH}& ={\rm Tr}_R [(-)^F
\bar q^{\bar L_0-{c_R\over 24}} e^{-\phi^Aq_A -\phi^0(L_0-{c_L\over 24})}]\cr
& = \left|Z_{top}(g_{top}={4\pi^2 \over \phi^0} ,
t^A = {\phi^A-\pi ip^A\over \phi^0})\right|^2 .}}
The first line is the OSV definition of the mixed partition function
and the second is the OSV relation to the square of the topological string partition function.

In conclusion we have rederived the OSV relation in all detail from an M-theory partition function on an $AdS_3\times S^2\times X$. In this picture the
factorization into holomorphic and antiholomorphic parts has a simple
origin as the contributions from M2-branes and anti-M2-branes.

\centerline{\bf Acknowledgement}
We are grateful to A. Adams, F. Denef, R. Dijkgraaf, D. Jafferis,
L. Huang, J. Maldacena, J. Marsano, H. Ooguri, C. Vafa,
and E. Verlinde for useful discussions.
This work is supported in part by DOE grant DE-FG02-91ER40654.

\appendix{A}{Ressumation of the GV formula}
In this appendix we rearrange the expression for the topological string partition function to express it in terms of the Gopakumar-Vafa invariants rather than
the degneracies $N_{q_A,j_L,j_R}$ of the irreducible representations.
Taking minus the log of the first product  in \llfs\ and summing over $j_L,j_R$, resumming and setting $g_{top}=-2\pi i\tau$, $t^A=y^A+{\tau \over 2}p^A$ gives
\eqn\llsfs{ \eqalign{F
 &=\sum_{q_A,n, m_L,j_Lj_R}  (-)^{2j_R+2j_L}n(2j_R+1)N_{q_A,j_L,j_R}\ln
(1-e^{-g_{top} (n+2m_L)} e^{2\pi iq_A t^A}
)\cr &=-\sum_{q_A,n, m_L,j_L,j_R,k}  (-)^{2j_R+2j_L}{n\over k}(2j_R+1)N_{q_A,j_L,j_R}e^{-kg_{top}
 (n+2m_L)} e^{2\pi ikq_A t^A}
\cr &=-\sum_{q_A,j_L,j_R,k}  (-)^{2j_R+2j_L}{1\over k}(2j_R+1)N_{q_A,j_L,j_R}{\sinh[(2 j_L+1)kg_{top}] \over \sinh^2[\half k g_{top}]  \sinh[kg_{top} ]}
 e^{2\pi ik q_A t^A}
\cr } }
In \gvtwo\ the Gopakumar-Vafa invariants $\alpha_{r,q_A}$ are defined by
\eqn\tyo{\sum_{r}  \alpha_{r,q_A}(-)^r(2\sinh {\theta \over 2} )^{2r}
=\sum_{j_L,j_R}  (-)^{2j_R+2j_L}(2j_R+1)N_{q_A,j_L,j_R}{\sinh[(2j_L+1)\theta]  \over \sinh[\theta]},}
so that
\eqn\jqa{F=\sum_{q_A,r,k}  {(-)^{r-1} \over k}\alpha_{r,q_A}(2\sinh{kg_{top}\over 2})^{2r-2}
 e^{2\pi ik q_A t^A}.}
 This agrees precisely with the expression for the topological string partition function given in \gvtwo.

\listrefs

\end
M2 WITH INTRINSIC SPIN, LONG STRINGS...
\eqn\reexp{ Z_{BH} (\tau={2\pi i\over \phi^0},
 y^A={\phi^A\over \phi^0}) = UZ^0_{sugra} (\tau={i\phi^0\over 2 \pi },
 y^A={i\phi^A }) = U |Z_{top}(g_{top}={\phi^0\over 2\pi },
t^A = {\phi^A-\pi ip^A\over \phi^0})|^2 }
To see this we recall from \gvone\  that for the contribution from isolated rational curves  \eqn\gvtop{
Z_{top}(g_{top},t^A) = \prod_{q_A,n} {1\over (1-e^{-g_{top}n- 2\pi
iq_A t^A})^{nN_Q}} } We shall restrict to the case where
$\tau=i\tau_2$ is purely imaginary
 and $y^A$ is real. Under the identification
\eqn\idtne{ g_{top} =  2\pi \tau_2,~~~~ t^A = y^A-\half\tau p^A } we
have
\eqn\zsuga{ Z_{sugra} (\tau, y^A) = |Z_{top}(g_{top},t^A)|^2 }
Using the attractor relation between the moduli and the
potentials
\eqn\idsatt{ g_{top} = {4\pi i\over X^0} = {4\pi^2\over \phi^0},
~~~~ t^A = {X^A\over X^0} = {\phi^A - \pi i p^A\over \phi^0} }
we recover the OSV relation

The elliptic genus of the dual $(0,4)$ CFT is
\eqn\elltr{ Z_{CFT}(\tau, y^A) = {\rm Tr}_R (-)^F q^{L_0-c_L/24}
\bar q^{\bar L_0-c_R/24} e^{2\pi i y^A q_A} }
where $q=e^{2\pi i\tau}$.
This is precisely the OSV mixed partition function
with the identification $\tau=i\phi^0/2\pi$ and
$y^A=i{\phi^A/ 2\pi}$. By spectral flow \elltr\ can be
written as a partition function summing over chiral primaries
on the right,
\eqn\elltns{ Z_{CFT}(\tau, y^A) = {\rm Tr}_{NS, \bar L_0
=J_R^3} (-)^F q^{L_0-c/24} e^{2\pi i y^A q_A} }

\newsec{String Interpretation}
In the previous section we computed the contribution to $Z_{sugra}$
from M2-branes wrapping a curve in the Calabi Yau and winding around the euclidean time circle. In this section we identify this circle with the
M-IIA compactification circle. The M2-branes then become string worldsheet instantons, and $Z_{sugra}$ identified with $|Z_{top}|^2$.

The $AdS_3$ metric in global coordinates is
\eqn\metr{{ds^2 \over R^2}= -cosh^2\chi dt^2+d\chi^2+sinh^2\chi d\phi^2,}
In these coordinates we have the Killing vectors
\eqn\dzk{L_0=\p_t+\p_\phi,~~~~\bar L_0=\p_t-\p_\phi.}
The M-theory partition function is computed on the Euclidean geometry obtained by $t\to i \tau$
\eqn\mestr{{ds^2 \over {\ell}^2}= cosh^2\chi d\tau^2+d\chi^2+sinh^2\chi d\phi^2,}
identified via a $\tau_2=g_{top}$ shift generated by
\eqn\jio{L_0=\p_\tau+\p_\phi.}
This implies
\eqn\kkl{\tau+\phi \sim \tau +\phi+g_{top}.}